\documentclass[%
 reprint,
superscriptaddress,
nofootinbib,
 amsmath,amssymb,
 aps,
 prd,
]{revtex4-2}

\usepackage{xcolor}
\usepackage{graphicx}%
\usepackage{dcolumn}%
\usepackage{bm}%
\usepackage{times}%
\usepackage{hyperref}%
\usepackage[normalem]{ulem}
\hypersetup{
  colorlinks=true,        %
  linkcolor=blue,         %
  citecolor=cyan,         %
  urlcolor=cyan            %
}

\newcommand{\ie}{i.e.,~}
\newcommand{\eg}{e.g.,~}

\begin{document}

\title{Probing neutron-star matter in the lab: similarities and differences between binary mergers and heavy-ion collisions}

\author{Elias R. Most}
\affiliation{Princeton Center for Theoretical Science, Princeton
  University, Princeton, NJ 08544, USA}
\affiliation{Princeton Gravity Initiative, Princeton University,
  Princeton, NJ 08544, USA}
\affiliation{School of Natural Sciences, Institute for Advanced Study,
  Princeton, NJ 08540, USA}

\author{Anton~Motornenko}
\affiliation{Institut f\"ur Theoretische Physik, Goethe Universit\"at,
  D-60438 Frankfurt am Main, Germany}
\affiliation{Frankfurt Institute for Advanced Studies, Giersch Science
  Center, D-60438 Frankfurt am Main, Germany}

\author{Jan Steinheimer}
\affiliation{Frankfurt Institute for Advanced
  Studies, Giersch Science Center, D-60438 Frankfurt am Main, Germany}

\author{Veronica Dexheimer}
\affiliation{Department of Physics, Kent State University, Kent, OH
  44243, USA}

\author{Matthias Hanauske}
\affiliation{Institut f\"ur Theoretische Physik, Goethe Universit\"at,
  D-60438 Frankfurt am Main, Germany}
\affiliation{Frankfurt Institute for Advanced Studies, Giersch Science
  Center, D-60438 Frankfurt am Main, Germany}

\author{Luciano Rezzolla}
\affiliation{Institut f\"ur Theoretische Physik, Goethe Universit\"at,
  D-60438 Frankfurt am Main, Germany}
\affiliation{Frankfurt Institute for Advanced Studies, Giersch Science
  Center, D-60438 Frankfurt am Main, Germany}
\affiliation{School of Mathematics, Trinity College, Dublin 2, Ireland}

\author{Horst Stoecker}
\affiliation{Institut f\"ur Theoretische Physik, Goethe Universit\"at,
  D-60438 Frankfurt am Main, Germany}
\affiliation{Frankfurt Institute for Advanced Studies, Giersch Science
  Center, D-60438 Frankfurt am Main, Germany}
\affiliation{GSI Helmholtzzentrum f\"ur Schwerionenforschung GmbH,
  D-64291 Darmstadt, Germany}

\date{\today}
\begin{abstract}
Binary neutron-star mergers and heavy-ion collisions are related
through the properties of the hot and dense nuclear matter formed
during these extreme events.  In particular, low-energy heavy-ion collisions
offer exciting prospects to recreate such {extreme} conditions in the laboratory.
However, it remains unexplored to what degree those collisions can actually
reproduce hot and dense matter formed in binary neutron star mergers.
As a way to understand similarities and differences between these
systems, we {discuss their geometry and }perform a direct numerical comparison of the thermodynamic
conditions probed in both collisions. To enable a direct comparison, we employ a 
finite-temperature equation of state  able to describe the entire high-energy phase diagram
of Quantum Chromodynamics. Putting side by side the evolution of both
systems, we find that laboratory heavy-ion collisions at the energy
range of $E_{\mathrm{lab}}=0.4 - 0.6\ A$ MeV probe
(thermodynamic) states of matter that are very similar to those created in binary neutron-star mergers.
These results can inform future low-energy heavy-ion collisions probing this regime.
\end{abstract}

\maketitle

\section{Introduction}
Quantum Chromodynamics~(QCD), the theory of strong interactions, predicts that matter at densities and temperatures found in binary neutron star mergers~{(BNSMs)} and relativistic heavy ion collisions~{(HICs)} consists of the same particles that obey the same interactions
despite drastic differences of system sizes.
Although hot dense matter is predicted to appear in a plethora of different states under different conditions~\cite{Rajagopal:1999cp, Alford:2002ng, Buballa:2003qv, Schafer:2003vz, Fukushima:2010bq}, in the laboratory it is only possible to measure the final hadronic states that are emitted from the late stages of HICs. This is what we refer to as nuclear matter, not only protons and neutrons that at low energy are clustered into nuclei, but also hyperons (that contain strange quarks), negative parity states, and  mesons. The life-time of HICs is so short, that full evolution conditions are problematic to reconstruct. Indirect observations of the Quark-Gluon plasma~(QGP) were done in high-energy collisions \cite{2000nucl.th...2042H,BRAHMS:2004adc,PHENIX:2004vcz,PHOBOS:2004zne,STAR:2005gfr}, though the details of such a state and the transitions that lead to it from nuclear matter and vice-versa are not yet known to great detail. Current state-of-the-art lattice QCD calculations of thermodynamic properties at vanishing baryon density (which corresponds to the fireball created in the highest-energy HICs, with the same number of particles and anti-particles for zero net-baryon density) suggest that the deconfinement transition from hadronic matter to the QGP appears smoothly~\cite{Borsanyi:2013bia,HotQCD:2014kol,Bazavov:2017dus}, i.e. there's no sharp boundary between the states. At finite, as well as high baryon densities, the phase structure is still rather unconstrained, and the existence of a first order phase transition, i.e. with a sharp boundary between the phases, is {not} ruled out. Verification of such a possibility is one of the top priority research tasks at several running and upcoming HIC facilities~\cite{Gazdzicki:2008kk,HADES:2009aat,Grebieszkow:2019yjd,CBM:2016kpk,Bzdak:2019pkr,HADES:2019auv,ALICE:2019nbs,STAR:2021iop,STAR:2021rls}.

Since the first gravitational detection of a {BNSM in 2017}, GW170817~\cite{LIGOScientific:2017vwq}, we have already employed the knowledge gained from mergers to improve our understanding of cold and dense nuclear matter by placing strong constraints on the neutron-star maximum mass \cite{Margalit:2017dij, Rezzolla:2017aly, Ruiz:2017due, Shibata:2019ctb} and typical radii and tidal deformabilities \cite{Annala:2017llu,Radice:2017lry, Bauswein:2017vtn, Most:2018hfd, LIGOScientific:2018cki,Raithel:2018ncd, De:2018uhw, Chatziioannou:2018vzf, Carson:2018xri,Dexheimer:2018dhb, Lim:2019som, Essick:2019ldf, Most:2020bba,Tan:2020ics,Tan:2021ahl,Tan:2021nat,Nathanail:2021tay}. 
In a similar manner, HICs have been suggested to provide complementary constraints.
Using flow studies, Ref. \cite{Danielewicz:2002pu} (see also \cite{Lynch:2009vc}) has extracted isotropic pressure constraints from low-energy HICs. These have recently been employed to (jointly with multi-messenger constraints) {derive} bounds on the cold dense matter EoS ~\cite{Huth:2021bsp}.
More fundamentally, this raises the question of what complementary information about hot and dense matter BNSMs can provide compared with HIC, given that both systems feature non-negligible temperatures. How similar is stellar nuclear matter to that produced in a HIC?
In particular, future BNSM events might open up the possibility to complement the experimental investigations of the short-lived microscopic nuclear matter created in HICs by detecting signals from the longer-lived macroscopic amount of (hot) nuclear matter formed in the merger. The main signatures would potentially be contained in the post-merger gravitational-wave emission \cite{Oechslin:2004yj,Bose2017,Most:2018eaw,Bauswein:2018bma,Most:2019onn,Weih:2019xvw,Liebling:2020dhf,Prakash:2021wpz,Tootle:2022pvd,Huang:2022mqp}, making this question a prime target for next-generation gravitational wave facilities \cite{Punturo:2010zz,Reitze:2019iox,Ackley:2020atn}.
On the HIC side, recent evidence of the formation of hot matter at several times saturation density in low-energy HICs was reported by the HADES collaboration \cite{HADES:2019auv}, conditions potentially resembling those present in a BNSM. However, in the absence of a direct and meaningful comparison between these two systems, it is far from clear exactly how similar the conditions probed in HICs and BNSMs really are.\\

Motivated by these claims, we aim to answer the following question: \textit{Can experiments such as \cite{HADES:2019auv} reproduce BNSM {geometry and thermodynamic} conditions, and what experimental ``settings'' are necessary?} \textit{How does gravity alter the outcome of macroscopic collisions, and what is actually meant by similar conditions in both cases?} To answer these questions, we perform a systematic and direct side-by-side comparison of geometry and thermodynamic properties in  HICs and BNSMs. This is done by means of relativistic hydrodynamics simulations of both systems. Crucially, we employ the same {realistic} microphysical EoS model, {describing} the entire high-energy QCD phase diagram, and the similar numerical methods to make the comparison as meaningful as possible. Introducing a quantitative figure of merit -- the entropy production during the collision-- to assess the similarity of the matter formed in both collisions, we derive bounds on the beam energies needed to probe BNSM-like conditions in laboratory HIC experiments.

\section{Methods}

In this work, we aim to compare {nuclear matter} in HICs and BNSMs by means of relativistic hydrodynamics simulations. 
To enable a meaningful comparison, we need to ensure two things. First, the microphysics needs to be the same in both cases, so that we can map the thermodynamic conditions present in one system to the other. Second, the numerical methods needs to be both reliable and comparable to ensure that effects such as shock heating are captured to the same degree in both simulations.

\subsection{Microphysical model}

The EoS used to describe both systems should be derived from a consistent model that is valid across the entire high-energy QCD phase diagram, covering the large range of temperatures and densities involved, in addition to describe different fractions of isospin and strangeness. Additionally,
such a model must be able to reproduce current constraints for compact stars and collider conditions (e.g., from Lattice QCD). More precisely, we consider constraints from cold compact stars~(\eg \cite{Raaijmakers:2021uju, Miller:2021qha, Fasano:2019zwm,Bauswein:2019skm, Capano:2019eae, Dietrich:2020efo, Landry:2020vaw,Al-Mamun:2020vzu}), properties of symmetric nuclear matter~
\cite{Zhou:2019omw,Zhang:2018vrx,Krastev:2018nwr, Reed:2021nqk, Tsang:2019jva,Kumar:2019oql, Sagun:2018sps, Li:2019tcx, Li:2021sxb, Biswas:2021yge,Huth:2021bsp, Biswal:2021mlf, Drischler:2021bup, Ferreira:2021pni},
and high-temperature QCD constraints at vanishing and low density, including a consistent description for chiral-symmetry restoration and quark deconfinement reproducing data
from lattice QCD \cite{Borsanyi:2013bia,HotQCD:2014kol,Bazavov:2017dus,Aarts:2017rrl}, perturbative QCD \cite{Fraga:2013qra,Haque:2014rua}, and high-energy collider experiments~\cite{Pratt:2015zsa}.

Addressing all of the above, we make use of the Chiral Mean Field~(CMF) model, which is based on the three-flavor chiral Lagrangian for hadronic matter first introduced in~\cite{Papazoglou:1998vr} and extended to describe neutron stars in
\cite{Dexheimer:2008ax}. Our version of the CMF model~\cite{Motornenko:2019arp} uniquely incorporates a full list of QCD degrees of freedom, including, besides protons and neutrons, hyperons, their parity partners, and the full list of hadronic resonances (strange and non-strange baryons and mesons) as found in the particle data book \cite{Workman:2022ynf}. In addition, the thermal contribution of deconfined quarks and gluons is added as in the PNJL approach \cite{Rischke:1991ke, Fukushima:2003fw, Ratti:2005jh,Steinheimer:2009hd,Zyla:2020zbs}. Together with the electrons this corresponds to the most complete set of QCD degrees of freedom available for the high density equation of state. A more detailed description on the CMF model and its contents can be found in e.g. \cite{Motornenko:2020yme}. In addition, a low density model that includes the description of nuclei was gradually matched to the CMF-EoS, for the BNSM simulations, below $10^{-2}~ n_{\rm sat}$. See Ref.~\cite{Schneider:2017tfi} for details on the matching and the low-density EoS.\\

Our version of the CMF model reproduces a crossover transition for deconfinement at finite and zero density (as determined by lattice QCD)~\cite{Motornenko:2018hjw,Motornenko:2020yme}, and providing a good description for hadrons in medium, nuclei, nuclear matter and neutron stars \cite{Motornenko:2019arp,Motornenko:2019lwh}. The latter includes reproducing ${M}>2 M_\odot$ stars and stars with radii within LIGO-Virgo and NICER allowed regions. The model produces a nuclear ground state with realistic properties: {saturation at (baryon number) density ${\rm n_{\rm sat}}=0.15 \ \mathrm{fm^{-3}}$}, binding energy per nucleon $E_0/B =-15.2$~MeV, symmetry
energy $S_0 = 31.9$~MeV, symmetry
energy slope $L = 57$~MeV, and incompressibility $K_0 = 267$~MeV. This comprehensive approach
allows to calculate the EoS of nuclear matter created in HICs and in BNSMs, without introducing additional ambiguities due to the use of (potentially inconsistent) different EoS in the two regimes. In addition, for BNSMs we {include a free gas of} electrons to the EoS to maintain electric charge neutrality. We calculate numerical tables of the temperature and density, as well as isospin dependence (e.g. difference between the amount of neutrons and protons) of the CMF-EoS, which are then implemented in the hydrodynamical models.

\begin{figure}[t!]
    \centering
    \includegraphics[width=.48\textwidth]{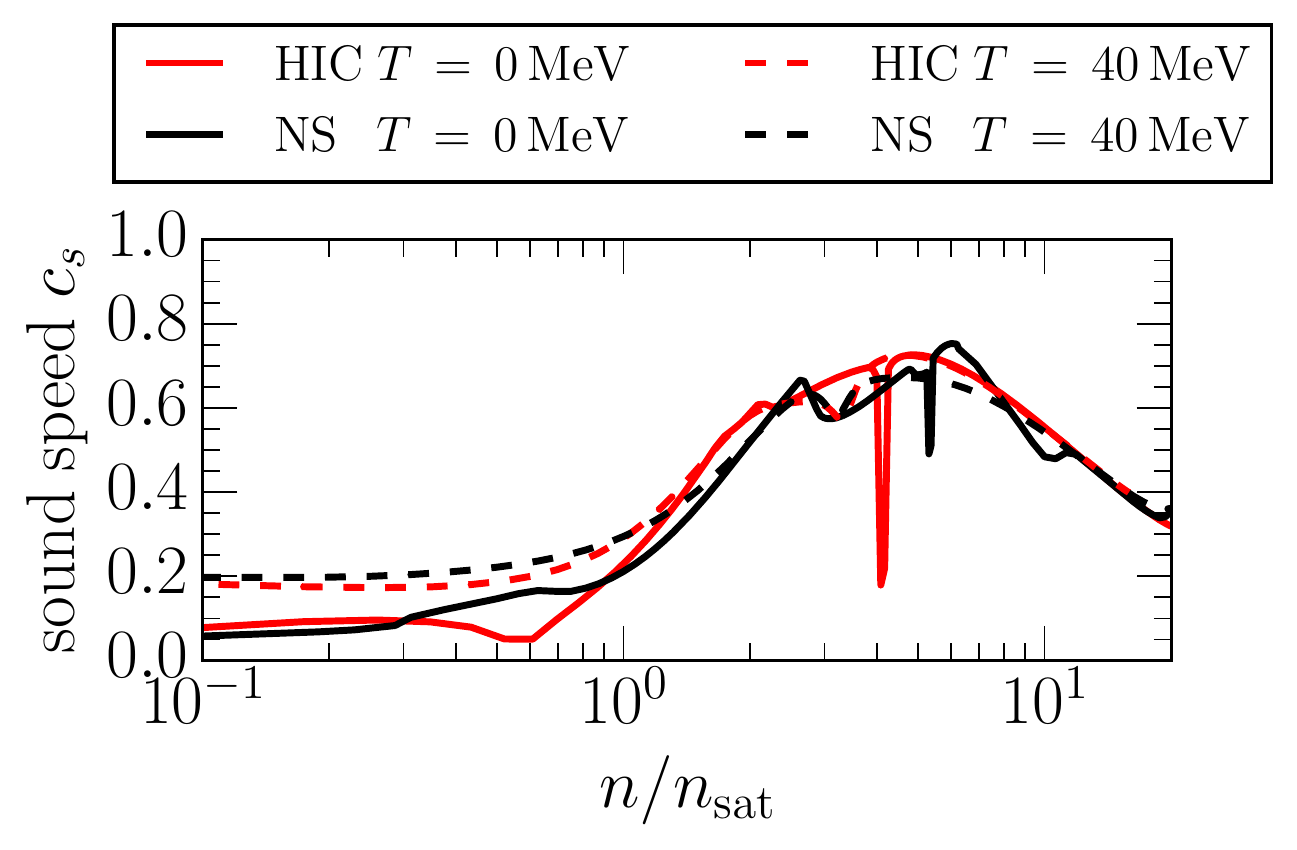}
    \caption{Isothermal sound speed, $c_s$, in the CMF equation of state for different scenarios. Black lines: Neutron star matter, including electrons in beta equilibrium and fulfilling charge neutrality. {{Red}} lines: Heavy ion collision matter with zero isospin and strangeness conservation. The solid lines correspond to the speed of sound at $T=0$ MeV and the dashed lines to $T=40$ MeV. A maximum in the speed of sound is clearly visible for all scenarios despite differences in the details due to the varying chemical composition. }
    \label{fig:0}
\end{figure}

In order to put the following simulation results into context of the EoS and to justify our assumption that HICs and BNSMs can be used complementary to study the QCD EoS, the speed of sound from the CMF model is presented in Fig.~\ref{fig:0} {(see also Refs~\cite{Tan:2021ahl,Altiparmak2022,Ecker2022}} for recent studies on the general properties of the sound speed in neutron stars). In particular, we show the isothermal speed of sound, $c_s$, as function of the baryon density at {some relevant} fixed temperatures of $T=0$ MeV and $T=40$ MeV. In both scenarios, we also compare the CMF results for either neutron star matter, including electrons in beta-equilibrium, as well as {HIC} matter which is isospin symmetric and where strangeness is conserved (number of strange particles matches the number of strange antiparticles). {The last two conditions are a consequence of the very short lived character of HICs.} Noticeably, the overall structure of the EoS is very similar for these two extreme cases: a clear maximum in the speed of sound is observed for 4-6 times saturation density (see also Ref.~\cite{Altiparmak:2022} for a more general discussion on the maximum of the sound speed in neutron stars). Both the decrease from the maximum at high densities and the oscillations are linked to the appearance of new degrees of freedom. %

While the details of the curves show clear differences due to the different chemical compositions of the systems, it is important that only the similarities allow us to study the properties of the QCD EoS in these very different microscopic vs. macroscopic scenarios.

\subsection{Numerical methods}
The other critical component of our study is the dynamical description of both BNSMs and HICs using relativistic hydrodynamics, which essentially provides conservation laws for the baryon current and the stress-energy tensor (see, e.g, \cite{Font:2000pp} for a review).
More specifically, we assume that the systems are (to lowest order) described as compressible perfect fluids. 
Although other hydrodynamic implementations, incorporating viscous effects or even a microscopic transport description of HICs are possible, it was shown that, if the same EoS is used, the HIC system dynamics and entropy production, as calculated with the hydrodynamical approach, are quite similar to the complementary predictions of non-equilibrium transport models~\cite{Kuttan:2022zno,Inghirami:2022afu}, see also \cite{Song:2007ux, Bozek:2009dw, Karpenko:2013wva, Shen:2014vra,Pang:2018zzo, Bemfica:2020zjp, Chabanov:2021dee, Pandya:2021ief,Most:2021rhr}).
Similarly, in the case of BNSM, viscous effects largely arise from modified Urca interactions \cite{Most:2022yhe,Alford:2017rxf}. While these can lead to small changes in the gravitational wave emission, their impact on the thermodynamics is subdominant. However, they might play a fundamental role in adjusting the isospin fraction of dense matter \cite{Most:2022yhe}, which we ignore in the comparison presented here.
In spite of that, perfect hydrodynamics is perfectly able to capture entropy production in the compressional regime of the flow by means of local Rankine-Hugoniot shock junction conditions \cite{rezzolla2013relativistic}. In fact, a recent study \cite{Most:2022yhe} has shown that (microphysical) viscous entropy production in BNSMs {is} of the order of $0.1/\rm baryon$ (in natural units).

The dynamical description of the evolution of both BNSMs and HICs is kept on the same footing by evolving in time the equations of relativistic hydrodynamics on a three-dimensional grid for both scenarios, but employing different numerical implementations. In the case of HICs, we use the Frankfurt \texttt{SHASTA} code \cite{Boris:1973tjt,Rischke:1995ir} with a uniform grid spacing $\Delta x=0.2\,{\rm fm}$ and time step $\Delta t=0.08\,{\rm fm}/c$.
{Our} HIC initial state consists of two drops of cold zero temperature nuclear matter colliding head-on with Lorentz-contracted Woods-Saxon density distributions, propagating towards each other with relativistic speed in the center-of-mass frame of the collision. For each energy, a near central collision of two gold nuclei (${\rm Au}$) is computed at fixed {offset} ``impact parameter'' $b=2\,{\rm fm}$ at lab energies of $E_{\rm lab}=450$ and $600\,A\,{\rm MeV}$ per nucleon, corresponding to those available to the HADES collaboration for low-energy HICs at GSI~\cite{HADES:2019auv}.
For our {BNSM} simulations, in addition to the equations of general-relativistic hydrodynamics \cite{rezzolla2013relativistic}, we need to solve Einstein's equations in the conformal Z4 formulation
\cite{Hilditch:2012fp, Bernuzzi:2009ex, Alic:2011gg, Alic:2013xsa}. The full set of equations is evolved using the \texttt{Frankfurt/IllinoisGRMHD} (\texttt{FIL}) code \cite{Etienne:2015cea,Most:2019kfe,Loffler:2011ay}. Making use of nested box-in-box mesh refinement \cite{Schnetter:2003rb}, our simulations use $7$ levels of refinement with the highest resolution of $\Delta x=250\,\rm m$ and outer box size of $1500\, \rm km$.  The adopted resolution has been shown to be sufficient within the context of this work \cite{Most:2018eaw,Most:2019kfe}. The initial conditions are two equal-mass with total masses of $2.6$ and $2.8\,M_\odot$ \cite{Gourgoulhon:2000nn}. 
We neglect spins in line with the assumption that most systems {are essentially} irrotational at merger \cite{Bildsten:1992my}.

\begin{figure*}[t!]
    \centering
    \includegraphics[width=\textwidth]{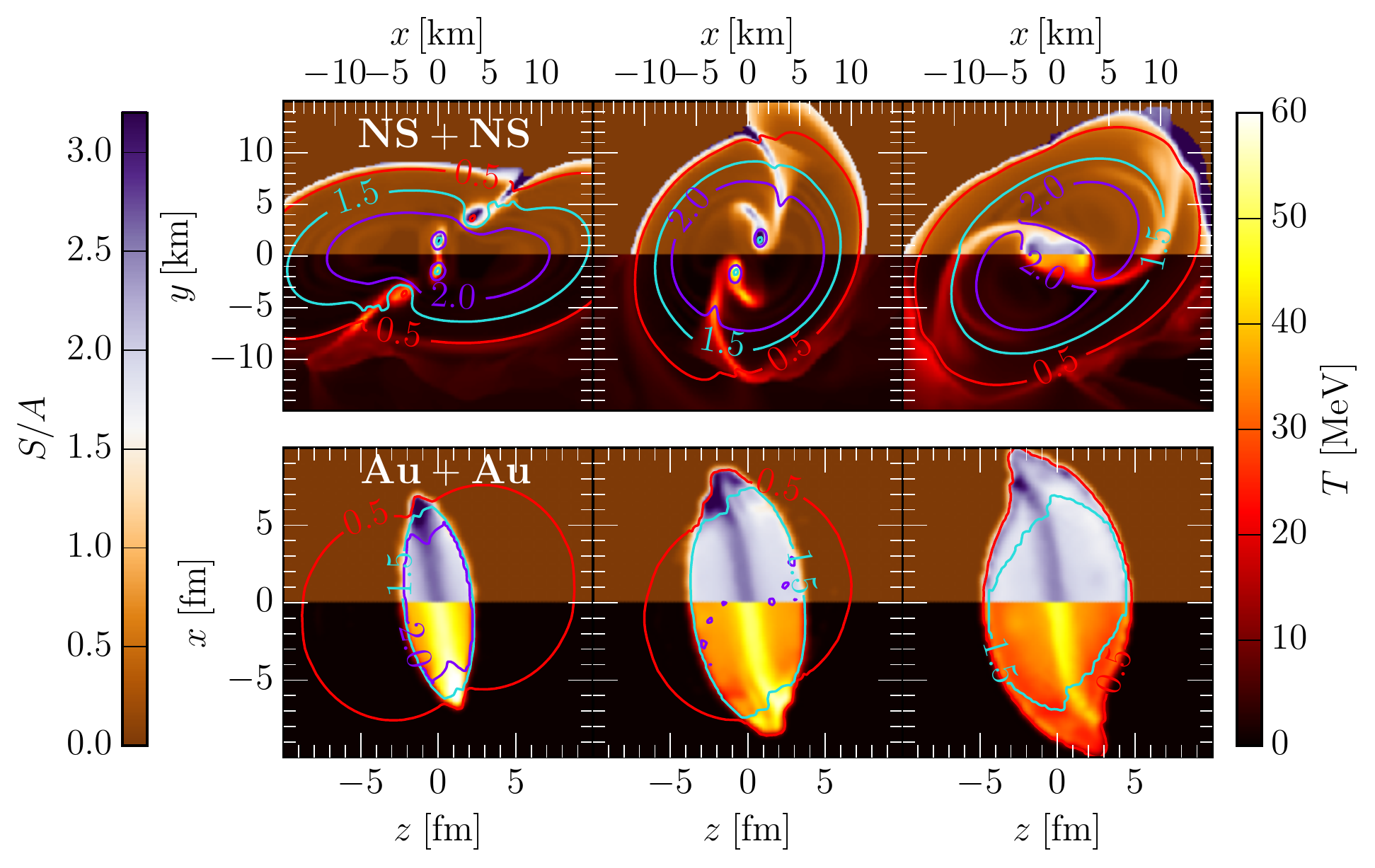}
    \caption{Distributions of entropy per baryon $S/A$ ({upper} colormaps)
      and temperature $T$ ({lower} colormaps) for a BNSM (NS+NS)
      with total mass $M_{\rm tot} = 2.8\, M_{\odot}$ ({top} panels) and a
      ${\rm Au+Au}$ HIC at $E_{\rm lab}=450\,A\,{\rm MeV}$ ({bottom} panels). Colored lines
      mark density contours in units of ${\rm n_{\rm sat}}$. The
      snapshots in different rows refer to $t=-2,\,0\,,+3\,{\rm ms}$  before and after
      merger for the BNSM, respectively, and to $t=-5\,,0\,,+5\,,\,{\rm fm}/c$ before and after the
      full overlap for the HIC.}
    \label{fig:1}
\end{figure*}

\section{Results}

In the following, we {{provide}} a comparison of low-energy HICs with BNSMs. Specifically, we {{focus}} on the {{geometry and}} thermodynamic properties of the matter probed in each system.
That is, because a full comparison of the flow structure for both systems is complicated by the fact that BNSMs have net angular momentum {{and}} are gravitationally bound. However, thermodynamic quantities in the local rest-frame of each fluid element in the collision remain meaningful.
 Since in the absence of physical viscosity the flow {{is}} isentropic, the entropies $S$ per baryon $A$ can serve as a meaningful tool to compare the flows in both cases.
Following along those lines, we proceed in three stages. First, we {provide} an overview of the collision dynamics in Sec. \ref{sec:sec1}. Second, in Sec. \ref{sec:sec2} we perform a detailed assessment of the entropy {{per baryon}} evolution in both cases. Finally, we perform a direct comparison of the phase-diagram coverage for different beam energies and BNS masses in Sec. \ref{sec:sec3}, establishing {{in which cases}} low-energy HICs can reproduce BNSM-like conditions.

\subsection{Overview of collision dynamics}
\label{sec:sec1}
In the following, we {{give}} an overview of the collision dynamics of both systems.
For a HIC, the two gold nuclei approach each other head-on along the $z$-direction, with relative velocities $v\gtrsim 0.5\,c$ and only a small offset $b$ along the transverse $x$-axis. This produces dense hot matter with the longest lifetime and highest compression (highest density) at a given beam energy. Once the two nuclei make contact, the cold nuclear matter in the center is rapidly heated and entropy is generated \cite{romatschke_romatschke_2019, Kuttan:2022zno}. Once both incoming nuclei are compressed into a single fireball, matter starts to rapidly expand along an isentropic trajectory until diluting so much that the hydrodynamic picture is no longer valid and freeze out occurs. In our simulations, this corresponds to cells at roughly $n
\sim \tfrac{1}{2}{\rm n_{\rm sat}}\approx 0.08\, \rm fm^{-3}\,$. At this point in the emitted nuclear matter will start to form clusters with important consequences for experimental studies of the EoS. 
Recent work has shown, that the maximum compression and entropy production from shock heating are rather insensitive to viscous corrections, when adopting  relatively low beam energies as considered in this work \cite{OmanaKuttan:2022the}. This allows us to safely neglect such corrections in the present comparison.
  
In the case of BNSs, the two stars are initially on a quasi-circular orbit, but the emission of gravitational waves causes the two stars to collide (see, \eg \cite{Baiotti:2016qnr} for a
review). Differently from a HIC, the collision is not head-on.  First, the merger remnant retains a significant fraction of angular momentum \cite{Hofmann:2016yih}. Second, tidal forces deform the neutron stars prior to merger, with small-scale turbulence induced in the shearing interface
between them (see, \eg \cite{Baiotti:2008ra}). During the merger, the two stars are compressed to a few times $n_{\rm sat}$ and heated considerably, leading to supersonic velocities and the formation of shocks. This causes a steep increase in temperature and a local production of entropy, similar to a HIC.

\begin{figure*}[t!]
    \centering
    \includegraphics[width=.95\textwidth]{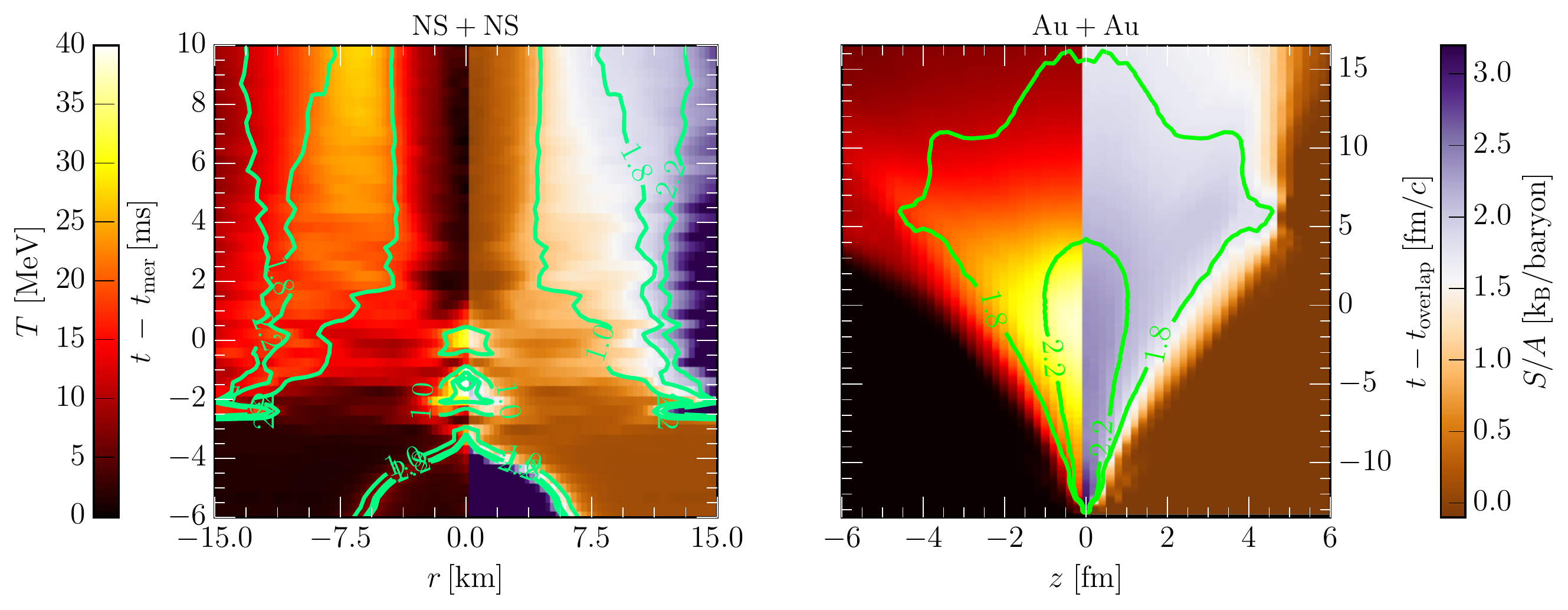}
    \caption{Dimensionally reduced spacetime diagrams for the evolution of the temperature and
      entropy {per baryon} relative to the same BNSM (left panel) and HIC (right
      panel) presented in Fig.~\ref{fig:1}. The green contours correspond
      to lines of constant entropy per baryon $S/A$; only regions with
      density above freeze-out, $n>\tfrac{1}{2}\,n_{\rm sat}$, are shown
      for the HIC.}
    \label{fig:2}
\end{figure*}

The differences in the {dynamics} between BNSMs and HICs are shown in {Fig.}~\ref{fig:1}, where we report the temperature $T$ (lower colormaps), the entropy per baryon $S/A$ (upper colormaps), and the density (isocontours) for a BNS collision ({top} panels) and a
${\rm Au+Au}$ collision ({bottom} panels). Note in the {top} row that, in spite of similar geometry, the entropy production in BNSMs is limited to a narrow spatial range at the interface of the two stars, where the densities probed are below $2\,n_{\rm sat}$. The precise structure of the merger remnant and, hence, the compression of layers, is governed by the strong gravitational fields present. Even some time into the collision, the heating and production of entropy is confined to the original collision interface. 

This is qualitatively different behavior from what happens in HICs, where, in the early phase of the collision, entropy production is also confined to a very thin ellipsoid in the narrow initial overlap of the two
nuclei. However, in the course of the reaction, the compression causes the entire gold nucleus (show as the contour with $n\gtrsim\tfrac{1}{2}\,n_{\rm sat}$) to heat up. Another important difference is that, after the {BNSM}, most of the resulting object is gravitationally bound, while in the HIC case, the
resulting remnant is an evanescent fireball of matter expanding isentropically at relativistic speeds. Finally, whereas in a BNSM, the overall rotation of the system and the conservation of the Bernoulli constant \cite{Hanauske:2016gia} leads to a redistribution of hot parts of the fluid, which ultimately settles down in a ring-like structure \cite{Kastaun:2016yaf, Hanauske:2016gia}, the hot fireball produced in a HIC cools rapidly, during the fast, isentropic expansion, with the central region always being at the highest temperature.

To summarize, even though the details of the dynamical evolution of both systems appear dramatically different, and the description of these can be rather technical, it is important to note that the bulk properties of the hot systems created are actually very similar. We discuss them in more detail in the following.

\subsection{Entropy evolution on micro- and macroscales}
\label{sec:sec2}

To {exploit} the similarities between BNSMs and HICs, we  compare the collision dynamics of the two systems by identifying the most important initial properties that lead to the same bulk entropy per baryon in the collision, namely, the beam energy in the case of HICs and the total stellar gravitational mass in BNSMs. To illustrate this behavior, we directly contrast the evolution of the entropy {per baryon} in the two different systems. The detailed comparison is shown in Fig.~\ref{fig:2} in terms of $1+1$ spacetime diagrams.  To
reduce the dimensionality of the collisions, we restrict to the equatorial $(x,y)$ plane for the BNSM and average out the azimuthal dependence, so that $r:=(x^2+y^2)^{\frac{1}{2}}$. In the case of the HIC, the whole transverse dependence in the $(x,y)$ plane is averaged out.

\begin{figure*}[t!]
    \centering \includegraphics[width=0.99\textwidth]{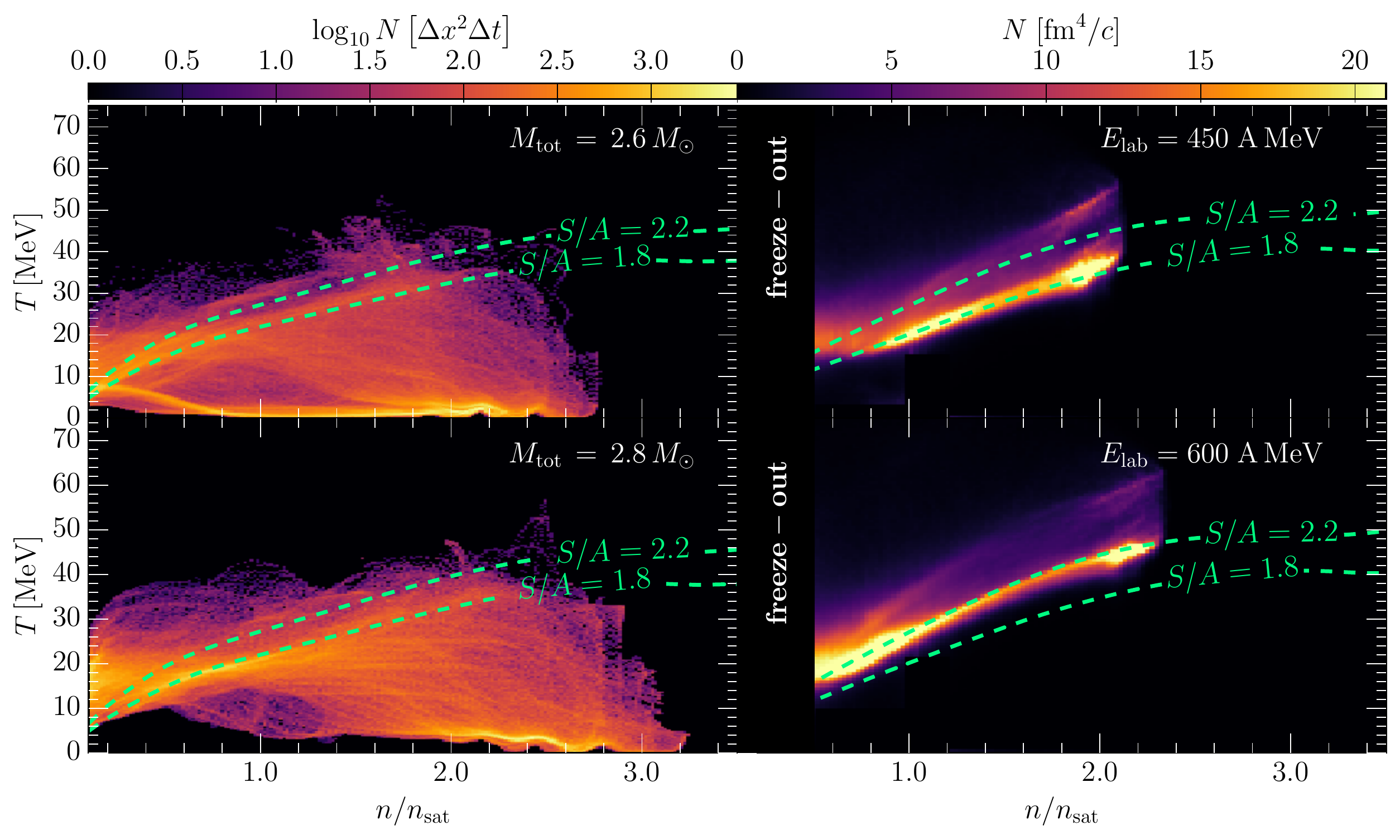}
    \caption{Regions of the QCD phase-diagram probed by two BNSMs
      with different masses (left panels) and by two HICs with different
      beam energies (right panels). The colorcode reports the number of
      cells $N$ in the various spacetimes having a given value of
      temperature and density. The green lines show contours of constant
      entropy per baryon. Only cells with
      density above freeze-out, $n>\tfrac{1}{2}\,n_{\rm sat}$, are shown for the {HICs}. In the BNSM case, we only consider the equatorial plane and normalize to arbitrary units.}
    \label{fig:13_ent}
\end{figure*}

The left panel of Fig.~\ref{fig:2} illustrates that before the collision
the temperatures and entropies of the stars are low. Tidal interactions
during the inspiral can lead to a mild heating of the outer layers of the
stars, but this is negligible as compared to the shock temperatures
reached in the merger \cite{Arras:2018fxj}. At the time of the merger,
\ie $t\simeq -3\, \rm ms$\footnote{The merger time is set when the
gravitational-wave amplitude has its first maximum \cite{Read:2013zra} a
few milliseconds after the stellar surfaces have touched
\cite{Takami:2014tva}.}, regions of very high temperature and entropy {per baryon} are
formed around $r \simeq 3 \,\rm km$. Note that the merger remnant
undergoes a significant thermodynamical evolution. The presence of high
angular momentum generates high shear flows despite of the fact that the
two stars were initially assumed irrotational. These shear flows
transport outwards the temperature and the entropy, leaving a
comparatively cold and dense core (see regions with $r \lesssim 3\,{\rm
  km}$) surrounded by a much hotter ring-like structure that remains
stationary in time \cite{Hanauske:2017oxo} ($4.0 \lesssim r\lesssim 7\,
\rm km$). The temperature and density further decrease when moving to the
outer regions of the remnant ($r\gtrsim 8\, {\rm km}$). A similar
behavior is shown by the evolution of the entropy distribution, which
exhibits a dense core with low entropy {per baryon}, surrounded by a hot ring with
$S/A \simeq 1-2$. We caution that the precise values of temperatures and entropies being probed might critically depend on physics around saturation \cite{Most:2021ktk}. A detailed investigation with a broader set of currently unavailable EoS will be necessary to more precisely estimate this error budget in both systems.

The spacetime diagram for a HIC (right panel of Fig.~\ref{fig:2}) shows
an ever increasing central-shock zone of high entropy {per baryon} and
high-temperature dense matter produced from the initial collision,
starting from $t\simeq -15\,{\rm fm}/c$ and up to the moment of full
overlap of the two nuclei. Subsequently, the interpenetration stage of
the two nuclei is over. Hence, the ellipsoid of arrested, hot,
shock-heated, and compressed matter can expand freely forwards and
backwards along the collision axes at $|z|>4~{\rm fm}$. The expansion of
matter is essentially isentropic, with average entropies of $S/A \approx
2.5$.

When comparing the two panels in Fig.~\ref{fig:2}, it is important to
note that there are striking analogies in the {thermodynamics properties of the} post-collision dynamics,
but also that these are confined mostly to the merger phase. This is
because the post-collision in a BNS is intrinsically different from that
in a HIC: in the former case, strong gravitational fields lead to a
remnant that is gravitationally bound and in a metastable equilibrium
\cite{Baiotti:2008ra}. By contrast, the hot and dense matter produced in
HICs is unbound and quickly expands into the surrounding vacuum.

\subsection{Comparison in the QCD phase diagram}
\label{sec:sec3}
Figure~\ref{fig:13_ent} shows the spatial \textit{and} temporal evolution
of these systems in the QCD phase diagram in terms of the temperature and
density of the various fluid cells. This is obtained by binning {into the variable $N$}, across
the whole evolution and for both BNSMs and HICs, all fluid elements
according to their temperature and density. {This means that a higher $N$ represents more regions that for longer timer have a given temperature and density.} As a simplification, for BNS
mergers we assume the equatorial plane dynamics to be representative of
the overall dynamics, thus performing a $2+1$ binning; for the HIC, the
full $3+1$ dynamics is used. For BNSs, we normalize to arbitrary units, since we are only interested in relative differences between populations of points in the phase diagram.  We show the results of two
distinct merger simulations differing in their total respective masses,
namely, for a binary with $M_{\rm tot} = 2.6\,M_\odot$ (top left panel)
and $2.8\,M_\odot$ (bottom left panel). {For} HICs, instead,
we show the results at two different beam (kinetic) energies, namely at
$E_{\rm lab}=450\, A\,{\rm MeV}$ (top right panel) and at $600\, A\,{\rm
  MeV}$ (bottom right {panel}).

The left panel of Fig.~\ref{fig:13_ent} shows a broad range of densities,
$n\lesssim 3.4\, {\rm n_{\rm sat}}$ and temperatures, $T\lesssim 40\,
{\rm MeV}$ covered by BNSMs. Two distinct regions appear in the
phase diagram during and after the merger: the first is at high
densities, $n> 2\, n_{\rm sat}$, and low temperatures, $T\lesssim\,
10\, {\rm MeV}$. This region corresponds to the central regions of the
initial stars and the core of the post-merger remnant. As we have seen in
Fig.~\ref{fig:2}, the neutron-star matter in this region does not undergo
shock heating, but remains cold and with low entropies. Indeed, the
oscillations seen at the lowest temperatures correspond to quadrupolar
post-merger oscillations of the gravitationally bound rotating remnant
\cite{Rezzolla:2016nxn}. This kind of matter is similar to cold stable neutron stars. 
The second region spanned by BNSMs in the
phase diagram corresponds, instead, to hot matter with $T\gtrsim10\,{\rm
  MeV}$ and isentropes of $S/A=\left[1.8,2.2\right]$, which were
previously identified with the hot ring in the discussion of
collision-shock dynamics (Fig.~\ref{fig:2}).  

The distributions in the left panels of Fig.~\ref{fig:13_ent} clearly indicate that the
lower mass binary populates regions with lower densities and
temperatures ($n \lesssim \, n_{\rm sat}$, $T\lesssim\, 10\, {\rm MeV}$),
which are essentially void in the case of the high-mass binary (see dark
region around $n \sim \tfrac{1}{2}\, n_{\rm sat}$). On the other hand,
low-density, high-temperature regions ($n \lesssim \, n_{\rm sat}$,
$T\lesssim\, 30\, {\rm MeV}$) are highly populated for the case of the
high-mass binary (see bright region around $n \lesssim 0.2\, n_{\rm
  sat}$). Although the BNM evolution cannot be described as isentropic, interestingly, for both BNS masses, the regions corresponding to similar isentropes of
$S/A\approx 2$ are populated. This range of entropies can be considered
as characteristic for the hot matter probed in BNSs mergers and
low-energy HICs. Because in our CMF model a significant quark fraction
can only build up at densities $n_{\rm b}\gtrsim 3\, n_{\rm sat}$ for the
low temperatures reached in both BNSs and HICs, $T\lesssim 80\,{\rm
  MeV}$, no deconfined matter is expected in these.

The right panels of Fig.~\ref{fig:13_ent} show the same as in the left
but for two HICs whose beam energies have been selected to provide
a distribution in phase diagram comparable to that of a BNSM, namely,
$E_{\rm lab}=450\, A\,{\rm MeV}$ and $600\, A\,{\rm MeV}$. Differently from the BNSMs, the evolution of HIC
remnants after the initial collision is an almost isentropic
expansion that populates the isentropes at $S/A~\sim~2$. Clearly,
different beam energies populate isentropes at lower/higher values of
$S/A$ (top/bottom panels). The rapid expansion from right to left along
the isentrope continues until matter becomes too dilute to maintain local
equilibrium and freezes out at $n \approx \tfrac{1}{2}\,n_{\rm sat}$.

The quite similar trajectories of the BNSM and the HICs in the QCD
phase diagram (concerning temperature, density, and entropy {per baryon}) may appear surprising at first sight. We recall that the nuclei used in HICs consist of an almost equal number of neutrons and protons (and among their parity partners), corresponding to nearly isospin-symmetric matter, \ie $Y_{\rm iso}\simeq -0.1$. Neutron-star matter is charge-neutral before the merger, consisting mainly of neutrons with a small admixture of protons, hyperons, parity partners, electrons, and muons in beta-equilibrium, \ie $[-0.5 \lesssim Y_{\rm iso} \lesssim -0.4]$ 
(see, \eg ~\cite{Most:2019onn,Most:2021ktk, Hammond:2021vtv, Perego:2019adq} for a more detailed discussion in the BNS-merger case). Nevertheless, these two different regimes are connected, and their relation is constrained at and around saturation density by the measurement of the {{(isospin)}} symmetry energy and its slope \cite{Tsang:2008fd}. It should be made clear that the knowledge on how the EoS changes with isospin is an essential ingredient for the comparison for the two systems and is exactly the reason why a consistent comparison can only be made on the basis of a single model.

\section{Conclusions}
In this work, we have set out to understand how well low-energy HICs are able to produce BNSM-like conditions \cite{HADES:2019auv}.
To this end, we have performed a set of numerical simulations modeling both low-energy HICs and BNSMs, so that we can directly compare the {{geometry and}} thermodynamic conditions present in each system. We have carefully designed the study to use the same microphysics, i.e., the same EoS, and comparable numerical methods to solve the relativistic hydrodynamics problem in flat and dynamically curved spacetime.
{{In}} order to mitigate a comparison of the very different flow structures in the presence of high net-angular momentum and strong gravity (for BNSMs), we have considered the local thermodynamic conditions probed. In particular, we have used the entropy production from shock heating at the initial impact as the figure of merit to meaningfully compare the two systems.
The main result of this comparion, is the use of
$S/A=\left[1.8,2.2\right]$ isentropes to construct a mapping between
gravitational masses of BNSs, \ie $M_{\rm tot} = 2.6-2.8\,M_{\odot}$, and
the beam energies of heavy-ion experiments conducted in laboratories, \ie
$E_{\rm lab}/A =\left[450, 600\right]\,{\rm MeV}$ \cite{CBM:2016kpk}. 
HICs at these beam energies are currently being investigated by the HADES experiment at the SIS18 accelerator of GSI.

This analysis can be extended in several ways. 
While the overall error of neglecting viscous effects {{is}} likely be small
in terms of the entropy production due to shock heating, a consistent study incorporating
microphysical viscosity in the HIC \cite{Monnai:2009ad,Song:2009rh,Bozek:2009dw,Dusling:2011fd,Noronha-Hostler:2013gga, Ryu:2015vwa,Ryu:2017qzn} and BNSM \cite{Alford:2017rxf,Most:2021zvc,Most:2022yhe} case would be desirable. 
In order to clarify uncertainties of the cold EoS on the entropy production in both systems, further studies will be required, once more EoSs covering the relevant phase space, such as ours, become widely available.
Our results present a significant step forward in the understanding of how
well low-energy HICs can probe BNSM-like conditions. We expect these to be 
particularly useful for the interpretation of hot-dense matter reported to be
formed in low-energy HICs \cite{HADES:2019auv}, 
even more so, should future gravitational wave detectors be able to provide independent constraints on the hot dense matter EoS \cite{Punturo:2010zz,Reitze:2019iox,Ackley:2020atn}.

\section*{Acknowledgements}
The authors thank M. Alford, T. Galatyuk, J. Noronha-Hostler, and
C. Raithel for insightful discussions and comments. ERM gratefully
acknowledges support as the John A. Wheeler fellow at the Princeton
Center for Theoretical Science, and through fellowships at the Princeton Gravity Initiative, and the
Institute for Advanced Study. AM acknowledges the Stern-Gerlach
Postdoctoral fellowship of the Stiftung Polytechnische Gesellschaft. JS
thank the Samson AG and the BMBF through the ErUM-Data project for
funding. VD acknowledges support from the National Science Foundation
under grants PHY1748621, NP3M PHY-2116686, and MUSES OAC-2103680. LR
acknowledges funding by the State of Hesse within the Research Cluster
ELEMENTS (Project ID 500/10.006), by the ERC Advanced Grant ``JETSET:
Launching, propagation and emission of relativistic jets from binary
mergers and across mass scales'' (Grant No. 884631), and by HGS-HIRe for
FAIR. HS acknowledges the Walter Greiner Gesellschaft zur F\"orderung der
physikalischen Grundlagenforschung e.V. through the Judah M. Eisenberg
Laureatus Chair at Goethe Universit\"at.

\bibliography{apssamp,non_inspire}%

\end{document}